\documentclass[10pt,twocolumn,article]{memoir}

\pdfoutput=1

\usepackage[keeplastbox]{flushend}

\setlrmarginsandblock{0.75in}{0.75in}{*}
\setulmarginsandblock{0.75in}{1in}{*}
\setcolsepandrule{0.33in}{0in}
\checkandfixthelayout

\counterwithout{section}{chapter}
\newcommand{\email}[1]{\normalsize{\texttt{#1}}}
\newcommand{\bigtablesize}{%
    \let\orignewcommand\newcommand%
    \let\newcommand\renewcommand%
    \makeatletter%
    \input{mem9.clo}%
    \makeatother%
    \let\newcommand\orignewcommand%
}

\usepackage{tikz}
\usepackage{forest}
\usepackage{framed}
\usepackage{cancel}
\usepackage{amssymb}
\usepackage{amsmath}
\usepackage{amsfonts}
\usepackage[shortlabels]{enumitem}

\newtheorem{demo}{Example}[section]
\newenvironment{demobis}[1]
    {\renewcommand{\thedemo}{\ref{#1}$'$}\addtocounter{demo}{-1}\begin{demo}}{\end{demo}}

\newcommand{\placeholder}{{-}}
\newcommand{\optional}[1]{\ulcorner\!#1\!\urcorner}
\newcommand{\selected}[1]{(\!(#1)\!)}
\newcommand{\textliteral}[1]{\mathtt{``#1"}}
\newcommand{\keyword}[1]{\mathsf{#1}}
\newcommand{\type}[1]{\mathsf{#1}}
\newcommand{\Wrap}[2]{#1{\left\{#2\right\}}}
\newcommand{\Opt}[1]{\Wrap{\type{Opt}}{#1}}
\newcommand{\Seq}[1]{\Wrap{\type{Seq}}{#1}}
\newcommand{\Rel}[1]{\Wrap{\type{Rel}}{#1}}
\newcommand{\Env}[2]{\Wrap{\type{Env}_{#1}}{#2}}
\newcommand{\Tuple}[1]{\langle #1 \rangle}
\newcommand{\Quotient}[2]{{#1}\Big/\raisebox{-1ex}{\!\small$#2$}}
\newcommand{\Void}{\type{Void}}
\newcommand{\Dept}{\type{Dept}}
\newcommand{\Emp}{\type{Emp}}
\newcommand{\Pos}{\type{Pos}}
\newcommand{\Text}{\type{Text}}
\newcommand{\Int}{\type{Int}}
\newcommand{\Bool}{\type{Bool}}
\newcommand{\Num}{\type{Num}}
\newcommand{\Department}{\keyword{department}}
\newcommand{\Employee}{\keyword{employee}}
\newcommand{\Name}{\keyword{name}}
\newcommand{\Position}{\keyword{position}}
\newcommand{\Salary}{\keyword{salary}}
\newcommand{\Manager}{\keyword{manager}}
\newcommand{\Subordinate}{\keyword{subordinate}}
\newcommand{\TopSalary}{\keyword{top\_salary}}
\newcommand{\Level}{\keyword{level}}
\newcommand{\Null}{\keyword{null}}
\newcommand{\Home}{\keyword{home}}
\newcommand{\True}{\keyword{true}}
\newcommand{\False}{\keyword{false}}
\newcommand{\Here}{\keyword{here}}
\newcommand{\Length}{\keyword{length}}
\newcommand{\Count}{\keyword{count}}
\newcommand{\Exists}{\keyword{exists}}
\newcommand{\Any}{\keyword{any}}
\newcommand{\All}{\keyword{all}}
\newcommand{\Sum}{\keyword{sum}}
\newcommand{\Max}{\keyword{max}}
\newcommand{\Min}{\keyword{min}}
\newcommand{\Mean}{\keyword{mean}}
\newcommand{\Select}{\keyword{select}}
\newcommand{\Filter}{\keyword{filter}}
\newcommand{\Define}{\keyword{define}}
\newcommand{\Size}{\keyword{size}}
\newcommand{\Total}{\keyword{total}}
\newcommand{\No}{\keyword{no}}
\newcommand{\Sort}{\keyword{sort}}
\newcommand{\Asc}{\keyword{asc}}
\newcommand{\Desc}{\keyword{desc}}
\newcommand{\Take}{\keyword{take}}

\newcommand{\Unique}{\keyword{unique}}
\newcommand{\Connect}{\keyword{connect}}
\newcommand{\Group}{\keyword{group}}
\newcommand{\Rollup}{\keyword{rollup}}
\newcommand{\Before}{\keyword{before}}
\newcommand{\Around}{\keyword{around}}
\newcommand{\Given}{\keyword{given}}
\newcommand{\Frame}{\keyword{frame}}
\newcommand{\To}{\boldsymbol{.}}
\newcommand{\Apply}{\!\boldsymbol{:}\!}
\newcommand{\As}{\Rightarrow}
\newcommand{\DEPARTMENT}{\keyword{D}}
\newcommand{\SALARY}{\keyword{S}}
\newcommand{\MEANSALARY}{\keyword{MS}}

\usetikzlibrary{arrows,arrows.meta,positioning}

\tikzset{
    entity diagram/.style={
        > = stealth',
        shorten > = 1pt,
        node distance = 1.6cm and 2.5cm,
        set/.style = {
            draw, ellipse, thick, font=\sffamily,
            minimum width=1.5cm, minimum height=.6cm,
            text height=1.5ex, text depth=.1ex},
        map/.style = {font=\small\sffamily}
    }
}

\tikzset{
    traverse/.style={
        ->,
        > = stealth',
        shorten < = 2pt,
        shorten > = 2pt
    }
}

\forestset{
    unfolded database/.style={
        void/.style = {
            draw, circle,
        },
        map/.style = {
            draw, rectangle, rounded corners=1mm,
            minimum width=2cm, minimum height=.5cm,
            text height=1.5ex, text depth=.25ex,
            font=\small\sffamily},
        box/.style = {
            draw, rectangle, rounded corners=1mm,
            minimum width=2cm, minimum height=.5cm,
            font=\small\sffamily},
        more/.style = {
            rectangle, minimum height=.5cm,
            edge path = {
                \noexpand
                \path [draw, -] (!u.parent anchor) -- +(0.15cm,0);
                \noexpand
                \path [draw, -, densely dotted] (!u.parent anchor) ++(0.15cm,0) -- +(0.15cm,0);
                \noexpand
                \path [draw, -] (!u.parent anchor) ++(0.1cm,0) -- +(0,-0.05cm);
                \noexpand
                \path [draw, -, densely dotted] (!u.parent anchor) ++(0.1cm,-0.05cm) -- +(0,-0.2cm);
            },
            l sep=0
        },
        and more/.style = {
            edge path = {
                \noexpand
                \path [\forestoption{edge}] (!u.parent anchor) -- +(0.1cm,0) |- (.child anchor);
                \noexpand
                \path [\forestoption{edge}] (!u.parent anchor |- .child anchor) ++(0.1cm,0) -- +(0,-0.05cm);
                \noexpand
                \path [draw, -, densely dotted] (!u.parent anchor |- .child anchor) ++(0.1cm,-0.05cm) -- +(0,-0.2cm);
            }
        },
        singular/.style = {
            edge={-}},
        optional/.style = {
            edge={-{Circle[open,length=3.5pt]}}},
        plural/.style = {
            edge={-{Straight Barb[reversed,length=3.5pt]}, shorten >=-0.5pt}},
        selected/.style = { thick, fill=lightgray!35 },
        selected edge/.style = { edge={thick} },
        for tree = {
            grow'=0,
            child anchor=west,
            parent anchor=east,
            anchor=west,
            calign=first,
            edge path={
                \noexpand
                \path [\forestoption{edge}] (!u.parent anchor) -- +(0.1cm,0) |- (.child anchor);
            },
        }
    }
}

\begin{document}

\title{Query Combinators}
\author{
    Clark C. Evans \\ \email{cce@clarkevans.com} \and
    Kyrylo Simonov \\ \email{xi@resolvent.net} \andnext
    Prometheus Research, LLC}
\date{Draft of \today}

\maketitle

\begin{abstract}
    We introduce Rabbit, a combinator-based query language.  Rabbit is designed
    to let data analysts and other accidental programmers query complex
    structured data.

    We combine the functional data model and the categorical semantics of
    computations to develop denotational semantics of database queries.  In
    Rabbit, a query is modeled as a Kleisli arrow for a monadic container
    determined by the query cardinality.  In this model, monadic composition
    can be used to navigate the database, while other \emph{query combinators}
    can aggregate, filter, sort and paginate data; construct compound data;
    connect self-referential data; and reorganize data with grouping and data
    cube operations.  A context-aware query model, with the input context
    represented as a como\-nadic container, can express query parameters and
    window functions.  Rabbit semantics enables pipeline notation, encouraging
    its users to construct database queries as a series of distinct steps, each
    individually crafted and tested.  We believe that Rabbit can serve as a
    practical tool for data analytics.
\end{abstract}

\section{Introduction}
\label{sec:introduction}

\emph{Combinators} are a popular approach to the design of compositional
domain-specific languages (DSLs). This approach views a DSL as an algebra of
self-con\-tained processing blocks, which either come from a set of predefined
atomic \emph{primitives} or are constructed from other blocks using block
combinators.

The combinator approach gives us a roadmap to design a database query language:

\begin{itemize}
\item
define the model of database queries;

\item
describe the set of primitive queries;

\item
describe the combinators for making composite queries.
\end{itemize}

To elaborate on this idea, we need some sample structured data.  Throughout
this paper, we use a simple database that contains just two classes of
entities: \emph{departments} and \emph{employees}.  Each department entity has
one attribute: \emph{name}.  Each employee entity has three attributes:
\emph{name}, \emph{position} and \emph{salary}.  Each employee belongs to a
department.  An employee may have a \emph{manager}, who is also an employee.

\begin{figure}
    \centering
    \begin{tikzpicture}[entity diagram]
        \node [set] (Dept) {Dept};
        \node [set] (Emp) [right=of Dept] {Emp};
        \node [set] (Text) [below=of Dept] {Text};
        \node [set] (Int) [below=of Emp] {Int};
        \draw [->] (Dept) to [bend right=15] node [map, right] {name} (Text);
        \draw [->] (Emp) to [bend right=15] node [map, right] {\;name} (Text);
        \draw [->] (Emp) to [bend left=15] node [map, right] {\;position} (Text);
        \draw [->] (Emp) to [bend left=15] node [map, right] {salary} (Int);
        \draw [->] (Emp) to node [map, above] {department} (Dept);
        \draw [->] (Emp) to [loop right] node [map, right] {manager} (Emp);
    \end{tikzpicture}
    \caption{Sample database}
    \label{fig:sample-schema}
\end{figure}
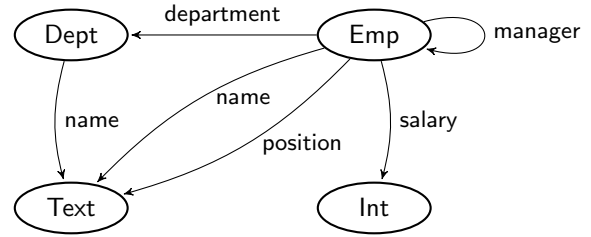

In Figure~\ref{fig:sample-schema}, the structure of the sample database is
visualized as a directed graph, with attributes and relationships (arcs)
connecting entity classes and attribute types (graph nodes).  This diagram may
suggest that we view attributes and relationships as functions with the given
types of input and output, for example
\begin{alignat*}{3}
    & \Department && : \Emp && \to \Dept, \\
    & \Name && : \Dept && \to \Text.
\end{alignat*}
This is known as the functional database model \cite{Kerschberg1976,
Sibley1977}.

This model provides us with a starting point on our combinator roadmap.
Indeed, a database query could be seen as a function; then, a set of primitive
queries is formed by all the attributes and relationships, while function
composition becomes a binary query combinator.  With these considerations, we
can write our first composite query.

\begin{demo}
    \label{ex:composite-query}
    Given an employee entity, show the name of their department.
    \begin{equation*}
        \Department\To\Name: \Emp \to \Text
    \end{equation*}
\end{demo}

In this example, $\Department\To\Name$ is a query written in Rabbit notation,
and $\Emp \to \Text$ is its signature.  The period (``$\To$'') denotes the
composition combinator, which is a polymorphic binary operator with a signature
\begin{equation*}
        \placeholder\,\To\,\placeholder : (A \to B,\; B \to C) \to (A \to C).
\end{equation*}

Even though this query model can express one database query, it does not seem
to be powerful enough to become the foundation of a query language.  What is
this model missing?

First, it is awkward that a query always demands an input.  It means that we
cannot express an input-free query like \emph{show a list of all
employees}.\footnote{We italicize \emph{business questions} that specify
$\keyword{database}$ $\keyword{queries}$.}

Further, although the relationships are bidirectional, the model only covers
one of their directions.  Indeed, we chose to represent the relationship
between departments and employees as a primitive with input $\Emp$ and output
$\Dept$.  However, we may just as well be interested in finding, \emph{for any
given department, the corresponding list of employees}.  It would be natural to
add a primitive for the opposite direction, but it cannot be encoded as a
function because its signature $\Dept \to \Emp$ would incorrectly imply that
there is exactly one employee per department.  Thus, the query model is unable
to express multivalued or \emph{plural} relationships.

The model also fails to capture the semantics of \emph{optional} attributes and
relationships.  Such is the relationship between employees and their managers,
which, according to Figure~\ref{fig:sample-schema}, should be encoded by a
primitive with signature $\Emp \to \Emp$.  But this signature implies that
every employee must have a manager, which is untrue.  Apparently, a pure
functional model is too restrictive to express the variety of relationships
between database entities.

This paper shows how to complete this query model and build a query language on
top of it.  It is organized as follows.

In Section~\ref{sec:cardinality}, we show how to represent optional and plural
relationships using the notion of query cardinality, which, following the
approach of categorical semantics of computations~\cite{Moggi1991}, determines
the monadic container for the query output.  This lets us establish a
compositional model of database queries.

In Section~\ref{sec:combinators}, we show how common data operations can be
expressed as query combinators.  Specifically, we describe combinators that
extract, aggregate, filter, sort and paginate data; construct compound data; and
connect self-referential data.

In Section~\ref{sec:quotients}, we show how grouping and data cube operations
can be implemented as combinators that reorganize the intrinsic hierarchical
structure of the database.

In Section~\ref{sec:context}, using the approach to the semantics of dataflow
programming~\cite{Uustalu2005}, we extend the query model to include a
comonadic query context, which allows us to express query parameters and window
functions.

In Section~\ref{sec:conclusion}, we summarize the query model and briefly
discuss some related work.

\section{Query Cardinality}
\label{sec:cardinality}

In Section~\ref{sec:introduction}, we suggested that a database query could be
modeled as a function.  However, this na\"{\i}ve model failed to represent
optional and plural relationships as well as queries lacking apparent input.
In this section, we resolve these issues by introducing the notion of
\emph{query cardinality}.

We found it difficult to model these two relationships:
\begin{enumerate}[(i)]
\item \label{itm:employee-to-manager}
\emph{An employee may have a manager.}
\item \label{itm:department-to-employee}
\emph{A department is staffed by a number of employees.}
\end{enumerate}
We were also puzzled on how to express input-free queries such as:
\begin{enumerate}[(i)]
\setcounter{enumi}{2}
\item \label{itm:employee-set}
\emph{Show a list of all employees.}
\end{enumerate}

We could attempt to represent optional and plural output values as instances of
the container types
\begin{equation*}
    \Opt{A} \quad \text{and} \quad \Seq{A},
\end{equation*}
where the \emph{option} container $\Opt{A}$ holds zero or one value of type
$A$, and the \emph{sequence} container $\Seq{A}$ holds an ordered list of
values of type~$A$.  Using these containers,
relationships~\ref{itm:employee-to-manager}
and~\ref{itm:department-to-employee} could be expressed as primitive queries
with signatures
\begin{alignat*}{3}
    & \Manager && : \Emp && \to \Opt{\Emp}, \\
    & \Employee && : \Dept && \to \Seq{\Emp}.
\end{alignat*}
Moreover, we could guess the output of query~\ref{itm:employee-set}.  Indeed,
\emph{a list of all employees} can only mean $\Seq{\Emp}$.

To describe the input of query~\ref{itm:employee-set}, we introduce a
\emph{singleton} type
\begin{equation*}
    \Void.
\end{equation*}
The type $\Void$ has a unique inhabitant ($\top:\Void$), and because there is
no freedom in choosing a value of this type, it can designate input that can
never affect the result of a query.  Using the singleton type, we can
express~\ref{itm:employee-set} as a \emph{class} primitive
\begin{equation*}
    \Employee : \Void \to \Seq{\Emp}.
\end{equation*}
Although both \ref{itm:department-to-employee} and \ref{itm:employee-set} are
denoted by the same name, we can still distinguish them by their input type.

Unfortunately, although containers let us represent optional and plural output,
they do not compose well.  For example, it is tempting to express \emph{for a
given employee, find their manager's salary} as a composition
\begin{equation} \label{eq:manager-to-salary}
    \Manager\To\Salary, \tag{$\star$}
\end{equation}
or \emph{show the names of all employees} as
\begin{equation} \label{eq:employee-to-name}
    \Employee\To\Name. \tag{$\star\star$}
\end{equation}
However, if we look at the signatures of the components
\begin{alignat*}{6}
    & \Manager && : \Emp && \to \Opt{\Emp}, \quad && \Salary && : \Emp && \to \Int, \\
    & \Employee && : \Void && \to \Seq{\Emp}, \quad && \Name && : \Emp && \to \Text,
\end{alignat*}
we see that their intermediate types do not agree, which means their
compositions are ill-formed.

A technique for composing queries can be found in the categorical semantics of
computational effects~\cite{Moggi1991}.  In this semantics, a program that maps
the input of type $A$ to the output of type $B$ is seen as a \emph{Kleisli
arrow} $A \to \Wrap{M}{B}$, where $M$ is a \emph{monad} that encapsulates the
program's effects.  Further, a sequential execution of programs $A \to
\Wrap{M}{B}$ and $B \to \Wrap{M}{C}$ is represented by their \emph{monadic
composition}, which is again a Kleisli arrow $A \to \Wrap{M}{C}$.

To utilize monadic composition, we distinguish the output type of a query from
the output container, which we call the query cardinality.  For example, we say
that query~\ref{itm:employee-to-manager} is an optional query from $\Emp$ to
$\Emp$, \ref{itm:department-to-employee} is a plural query from $\Dept$ to
$\Emp$, and \ref{itm:employee-set} is a plural query from $\Void$ to $\Emp$.
Then, any two queries should compose, regardless of their cardinalities, so
long as they have compatible intermediate types; furthermore, the least upper
bound of their cardinalities is the cardinality of their composition.

Specifically, given two queries
\begin{equation*}
    p : A \to \Wrap{M_1}{B}, \qquad q : B \to \Wrap{M_2}{C}
\end{equation*}
we first promote their output to a common cardinality
\begin{equation*}
    M = M_1 \sqcup M_2,
\end{equation*}
and then use the monadic composition combinator
\begin{equation*}
    \placeholder\,\To\,\placeholder: (A \to \Wrap{M}{B},\; B \to \Wrap{M}{C}) \to (A \to \Wrap{M}{C}).
\end{equation*}
to construct
\begin{equation*}
    p\,\To\,q : A \to \Wrap{M}{C}.
\end{equation*}
Using this rule, we can justify the queries (\ref{eq:manager-to-salary}) and
(\ref{eq:employee-to-name}) and give them signatures
\begin{alignat*}{3}
    & \Manager\To\Salary && : \Emp && \to \Opt{\Int}, \\
    & \Employee\To\Name && : \Void && \to \Seq{\Text}.
\end{alignat*}

Let us work out the details.  Query cardinalities are ordered with respect to
inclusions
\begin{equation*}
    A \sqsubseteq \Opt{A} \sqsubseteq \Seq{A},
\end{equation*}
which, using the notation for container instances
\begin{equation*}
    \bot,\; \optional{a} : \Opt{A}, \qquad [a_1,\ldots,a_n] : \Seq{A},
\end{equation*}
are defined by
\begin{alignat*}{5}
    & \; && \bot && : \Opt{A} && \longmapsto [\;] && : \Seq{A}, \\
    & a : A \longmapsto\ && \optional{a} && : \Opt{A} && \longmapsto [a] && : \Seq{A}.
\end{alignat*}
This order lets us, whenever necessary, promote any query $A\to\Wrap{M}{B}$ to
a query $A\to\Wrap{M'}{B}$ with a greater cardinality $M' \sqsupseteq M$.

Monadic composition for the option and sequence containers is well known.  For
optional queries
\begin{equation*}
    p : A \to \Opt{B}, \qquad q : B \to \Opt{C},
\end{equation*}
it is defined by
\begin{align*}
    & p\,\To\,q : A \to \Opt{C}, \\
    & p\,\To\,q : a \mapsto \begin{cases}
        \optional{c} & (p(a)=\optional{\,b},\; q(b)=\optional{c}), \\
        \bot & (\text{otherwise}).
    \end{cases}
\end{align*}
For plural queries
\begin{equation*}
    p : A \to \Seq{B}, \qquad q : B \to \Seq{C},
\end{equation*}
the sequence $(p\,\To\,q)(a)$ is calculated by applying $p$ to $a$
\begin{equation*}
    a \overset{p}{\longmapsto} [b_1, b_2, \ldots],
\end{equation*}
then applying $q$ to every element of $p(a)$
\begin{equation*}
    [b_1, b_2, \ldots]
    \overset{[q]}{\longmapsto}
    [[c^{1}_{1}, c^{2}_{1}, \ldots], [c^{1}_{2}, c^{2}_{2}, \ldots], \ldots],
\end{equation*}
and finally merging the nested sequences
\begin{equation*}
    [[c^{1}_{1}, c^{2}_{1}, \ldots], [c^{1}_{2}, c^{2}_{2}, \ldots], \ldots]
    \overset{\cancel{\,[\;]\,}}{\longmapsto}
    [c^{1}_{1}, c^{2}_{1}, \ldots, c^{1}_{2}, c^{2}_{2}, \ldots].
\end{equation*}

At last, we are ready to present the design of a com\-binator-based query
language.

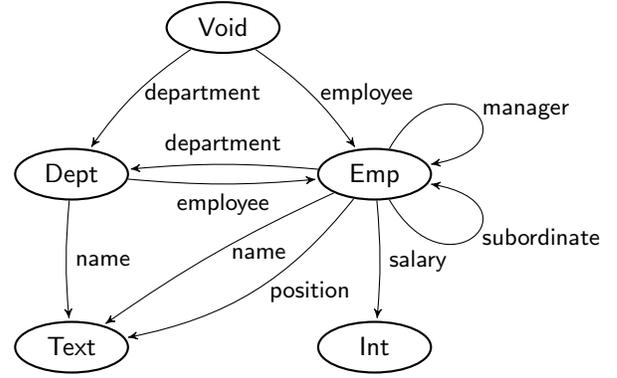
\begin{figure}
    \centering
    \begin{tikzpicture}[entity diagram]
        \node [set] (Dept) {Dept};
        \node [set] (Emp) [right=of Dept] {Emp};
        \coordinate (BetweenDeptAndEmp) at ($(Dept)!0.5!(Emp)$);
        \node [set] (Void) [above=of BetweenDeptAndEmp]{Void};
        \node [set] (Text) [below=of Dept] {Text};
        \node [set] (Int) [below=of Emp] {Int};
        \draw [->] (Void) to [bend right=10] node [map, right] {department} (Dept);
        \draw [->] (Void) to [bend left=10] node [map, right] {employee} (Emp);
        \draw [->] (Dept) to [bend right=5] node [map, right] {name} (Text);
        \draw [->] (Dept) to [bend right=5] node [map, below] {employee} (Emp);
        \draw [->] (Emp) to [bend right=5] node [map, right] {\;name} (Text);
        \draw [->] (Emp) to [bend left=20] node [map, right] {\;position} (Text);
        \draw [->] (Emp) to [bend left=5] node [map, right] {salary} (Int);
        \draw [->] (Emp) to [bend right=5] node [map, above] {department} (Dept);
        \draw [->] (Emp) to [loop right,out=60,in=10,looseness=7] node [map, right] {manager} (Emp);
        \draw [->] (Emp) to [loop right,out=300,in=350,looseness=7] node [map, right] {subordinate} (Emp);
    \end{tikzpicture}
    \caption{Database schema in folded form}
    \label{fig:folded-form}
\end{figure}

\textbf{Query model.} A database query is characterized by its input type $A$,
its output type $B$ and its cardinality $M$, and can be represented as a
function of the form
\begin{equation*}
    p : A \to M\{B\},
\end{equation*}
where $M\{B\}$ is one of $B$, $\Opt{B}$ or $\Seq{B}$; the respective queries
are called singular, optional or plural.

\textbf{Primitives.} The set of primitives includes classes
\begin{alignat*}{3}
    & \Department && : \Void && \to \Seq{\Dept}, \\
    & \Employee && : \Void && \to \Seq{\Emp};
\end{alignat*}
attributes
\begin{alignat*}{6}
    & \Name && : \Dept && \to \Text, \qquad
    && \Name && : \Emp && \to \Text, \\
    & \Position && : \Emp && \to \Text, \qquad
    && \Salary && : \Emp && \to \Int;
\end{alignat*}
and relationships
\begin{alignat*}{3}
    & \Department && : \Emp && \to \Dept, \\
    & \Employee && : \Dept && \to \Seq{\Emp}, \\
    & \Manager && : \Emp && \to \Opt{\Emp}, \\
    & \Subordinate && : \Emp && \to \Seq{\Emp}.
\end{alignat*}

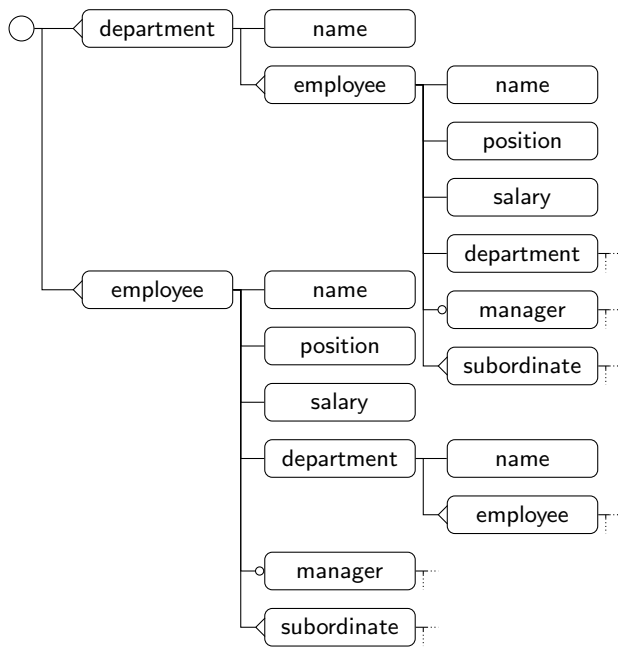
\begin{figure}
    \centering
    \begin{forest}
        unfolded database
        [,void
            [$\Department$,map,plural
                [$\Name$,map,singular]
                [$\Employee$,map,plural
                    [$\Name$,map,singular]
                    [$\Position$,map,singular]
                    [$\Salary$,map,singular]
                    [$\Department$,map,singular
                        [,more]]
                    [$\Manager$,map,optional
                        [,more]]
                    [$\Subordinate$,map,plural
                        [,more]]
                    [,phantom]]]
            [$\Employee$,map,plural
                [$\Name$,map,singular]
                [$\Position$,map,singular]
                [$\Salary$,map,singular]
                [$\Department$,map,singular
                    [$\Name$,map,singular]
                    [$\Employee$,map,plural
                        [,more]]]
                [$\Manager$,map,optional
                    [,more]]
                [$\Subordinate$,map,plural
                    [,more]]]]
    \end{forest}
    \caption{Database schema in unfolded form}
    \label{fig:unfolded-form}
\end{figure}

Recall that the original, incomplete set of primitives was obtained from the
schema graph in Figure~\ref{fig:sample-schema}.  To reflect the full set of
primitives, we should add the $\Void$ node and the remaining arcs (see
Figure~\ref{fig:folded-form}).  Furthermore, we can transform the schema graph
into an (infinite) tree by unfolding it starting from the $\Void$ node (see
Figure~\ref{fig:unfolded-form}).  The unfolded tree represents the functional
database in a \emph{universal hierarchical form}.

\textbf{Combinators.} The composition combinator sends two queries
\begin{equation*}
    p : A \to M_1\{B\}, \qquad
    q : B \to M_2\{C\}
\end{equation*}
to their composition
\begin{equation*}
    p\,\To\,q : A \to M\{C\} \qquad (M = M_1 \sqcup M_2).
\end{equation*}

Other common combinators are listed in Table~\ref{tab:common-combinators}.

\section{Query Combinators}
\label{sec:combinators}

In this section, we show how the query model defined in
Section~\ref{sec:cardinality} can support a wide range of operations on data.

\subsection*{Extracting Data}

By traversing the tree of Figure~\ref{fig:unfolded-form}, we can extract data
from the database.

\begin{demo}
    \label{ex:department-name}
    Show the name of each department.
    \begin{equation*}
        \Department\To\Name
    \end{equation*}
\end{demo}

This example is constructed by descending through nodes $\Department$ and
$\Name$, which represent primitives
\begin{alignat*}{3}
    & \Department && : \Void && \to \Seq{\Dept}, \\
    & \Name && : \Dept && \to \Text.
\end{alignat*}
The composition of the primitives inherits the input of the first component and
the output of the second component.  Since one of the components is plural, the
composition is also plural, which gives it a signature
\begin{equation*}
    \Department\To\Name : \Void \to \Seq{\Text}.
\end{equation*}

\begin{demo}
    \label{ex:department-employee-name}
    For each department, show the name of each employee.
    \begin{equation*}
        \Department\To\Employee\To\Name
    \end{equation*}
\end{demo}

This example takes a path through
\begin{alignat*}{3}
    & \Department && : \Void && \to \Seq{\Dept}, \\
    & \Employee && : \Dept && \to \Seq{\Emp}, \\
    & \Name && : \Emp && \to \Text
\end{alignat*}
to construct a query
\begin{equation*}
    \Department\To\Employee\To\Name : \Void \to \Seq{\Text}.
\end{equation*}

This query produces a list of employee names.  Since each employee belongs to
exactly one department, the list should contain the name of every employee.
The order in which the names appear in the output depends on the intrinsic
order of the $\Department$ and $\Employee$ primitives, but, in any case,
employees within the same department will be coupled together.

The same collection of names, although not necessarily in the same order, is
produced by the following example.

\begin{demo}
    \label{ex:employee-name}
    Show the name of each employee.
    \begin{equation*}
        \Employee\To\Name
    \end{equation*}
\end{demo}

On the other hand, the next example is very different from the apparently
similar Example~\ref{ex:department-name}.

\begin{demo}
    \label{ex:employee-department-name}
    For each employee, show the name of their department.
    \begin{equation*}
        \Employee\To\Department\To\Name
    \end{equation*}
\end{demo}

Here, we should see a list of department names, but each name will appear as
many times as there are employees in the corresponding department.

\begin{demo}
    \label{ex:employee-position}
    Show the position of each employee.
    \begin{equation*}
        \Employee\To\Position
    \end{equation*}
\end{demo}

Similarly, $\Employee\To\Position$ will output duplicate position titles.  We
will see how to produce a list of \emph{unique} positions in
Section~\ref{sec:quotients}.

\begin{demo}
    \label{ex:employee}
    Show all employees.
    \begin{equation*}
        \Employee
    \end{equation*}
\end{demo}

This example emits a sequence of employee entities, which, in practice, could
be represented as records with employee attributes.

\providecommand{\bigtablesize}{\relax}
\begin{table}
    \begin{framed}
    \bigtablesize
    \textbf{Identity and constants}
    \begin{alignat*}{3}
        & \Here && : A \to A && = a \mapsto a \\
        & 150000 && : A \to \Int && = a \mapsto 150000 \\
        & \Home && : A \to \Void && = a \mapsto \top \\
        & \Null && : A \to \Opt{B} && = a \mapsto \bot
    \end{alignat*}
    \textbf{Some scalar combinators}
    \begin{alignat*}{3}
        & {=},{\ne} && : (A \to B, A \to B) && \to (A \to \Bool) \\
        & {<},{\le},{>},{\ge} && : (A \to B, A \to B) && \to (A \to \Bool) \\
        & {\&},{|} && : (A \to \Bool, A \to \Bool) && \to (A \to \Bool) \\
        & {+},{-} && : (A \to \Int, A \to \Int) && \to (A \to \Int) \\
        & \Length && : (A \to \Text) && \to (A \to \Int)
    \end{alignat*}
    \textbf{Aggregate combinators}
    \begin{alignat*}{3}
        & \Count && : (A \to \Seq{B}) && \to (A \to \Int) \\
        & \Exists && : (A \to \Seq{B}) && \to (A \to \Bool) \\
        & \Any, \All && : (A \to \Seq{\Bool}) && \to (A \to \Bool) \\
        & \Sum && : (A \to \Seq{\Int}) && \to (A \to \Int) \\
        & \Max, \Min && : (A \to \Seq{\Int}) && \to (A \to \Opt{\Int})
    \end{alignat*}
    \textbf{Sequence transformers}
    \begin{alignat*}{4}
        & \Filter && : (&& A \to \Seq{B}, B \to \Bool) && \to (A \to \Seq{B}) \\
        & \Sort && : (&& A \to \Seq{B}, && \\
        & && && B \to C_1, \ldots, B \to C_n) && \to (A \to \Seq{B}) \\
        & \Take && : (&& A \to \Seq{B}, A \to \Int) && \to (A \to \Seq{B}) \\
        & \Unique : (A \to \Seq{B})\hidewidth && && && \to (A \to \Seq{B})
    \end{alignat*}
    \textbf{Selector and modifiers}
    \begin{alignat*}{3}
        & \Select & : (& A \to \Wrap{M}{B}, && \\
        & && B \to \Wrap{M_1}{C_1}, \ldots, B \to \Wrap{M_n}{C_n}) && \\
        \span\span\span\span \to (A \to \Wrap{M}{\Tuple{\Wrap{M_1}{C_1}, \ldots, \Wrap{M_n}{C_n}}}) & \\
        & \Define & : (& A \to \Wrap{M}{B}, B \to T) \to (A \to \Wrap{M}{B}) && \\
        & \Asc, \Desc\; & : (& A \to B) \to (A \to B_\lessgtr) &&
    \end{alignat*}
    \textbf{Hierarchical connector}
    \begin{equation*}
        \Connect : (A \to \Opt{A}) \to (A \to \Seq{A})
    \end{equation*}
    \textbf{Grouping}
    \begin{alignat*}{3}
        & \Group && : ( A \to \Seq{B}, B \to C_1, \ldots, B \to C_n) && \\
        \span\span\span\span \to (A \to \Seq{\Tuple{C_1, \ldots, C_n, \Seq{B}}}) & \\
        & \Rollup && : ( A \to \Seq{B}, B \to C_1, \ldots, B \to C_n) && \\
        \span\span\span\span \; \to (A \to \Seq{\Tuple{\Opt{C_1}, \ldots, \Opt{C_n}, \Seq{B}}}) &
    \end{alignat*}
    \textbf{Context primitives and combinators}
    \begin{align*}
        & \Frame : (\Rel{A} \to \Wrap{M}{B}) \to (A \to \Wrap{M}{B}) \\
        & \Before, \Around : \Rel{A} \to \Seq{A} \\
        & \Given : (\Env{T}{A} \!\to\! \Wrap{M}{B}, A \!\to\! T) \to (A \!\to\! \Wrap{M}{B}) \\
        & \textit{\textsf{PARAM}} : \Env{T}{A} \to T
    \end{align*}
    \vspace*{-\bigskipamount}
    \end{framed}
    \caption{Some primitives and combinators}
    \label{tab:common-combinators}
\end{table}

\subsection*{Summarizing Data}

Let us show how the extracted data can be summarized.

\begin{demo}
    \label{ex:count-department}
    Show the number of departments.
    \begin{equation*}
        \Count(\Department)
    \end{equation*}
\end{demo}

This query produces a single number, so that its signature is
\begin{equation*}
    \Count(\Department) : \Void \to \Int.
\end{equation*}
It is constructed by applying the $\Count$ combinator to a query that generates
\emph{a list of all departments}
\begin{equation*}
    \Department : \Void \to \Seq{\Dept}.
\end{equation*}
Comparing the signatures of these two queries, we can derive the signature of
the $\Count$ combinator, in this specific case
\begin{equation*}
    (\Void \to \Seq{\Dept}) \to (\Void \to \Int),
\end{equation*}
and, in general
\begin{equation*}
    \Count: (A \to \Seq{B}) \to (A \to \Int).
\end{equation*}
In other words, the $\Count$ combinator transforms any sequence-valued query
to an integer-valued query.  It is implemented by lifting the function that
computes the length of a sequence
\begin{equation*}
    |-| : \Seq{A} \to \Int
\end{equation*}
to a query combinator
\begin{equation*}
    \Count(q) = a \mapsto |q(a)|.
\end{equation*}

Unary combinators that transform a plural query to a singular (or optional)
query are called \emph{aggregate} combinators.

\begin{demo}
    \label{ex:max-employee-salary}
    What is the highest employee salary?
    \begin{equation*}
        \Max(\Employee\To\Salary)
    \end{equation*}
\end{demo}

In this example, we extract the relevant data with
\begin{equation*}
    \Employee\To\Salary : \Void \to \Seq{\Int}
\end{equation*}
and summarize it using the $\Max$ aggregate
\begin{equation*}
    \Max(\Employee\To\Salary) : \Void \to \Opt{\Int}.
\end{equation*}
This query is optional since it produces no output when the database contains
no employees.

\begin{demo}
    \label{ex:department-count-employee}
    For each department, show the number of employees.
    \begin{equation*}
        \Department\To\Count(\Employee)
    \end{equation*}
\end{demo}

In this example, we transform a plural relationship, \emph{all employees in the
given department}
\begin{equation*}
    \Employee : \Dept \to \Seq{\Emp}
\end{equation*}
to a calculated attribute, \emph{the number of employees in the given
department}
\begin{equation*}
    \Count(\Employee) : \Dept \to \Int.
\end{equation*}
Then we attach it to
\begin{equation*}
    \Department : \Void \to \Seq{\Dept}
\end{equation*}
to get \emph{the number of employees in each department}
\begin{equation*}
    \Department\To\Count(\Employee) : \Void \to \Seq{\Int}.
\end{equation*}

Applying the combinator $\Max$ to the query above, we answer the following
question.

\begin{demo}
    \label{ex:max-department-count-employee}
    How many employees are in the largest department?
    \begin{equation*}
        \Max(\Department\To\Count(\Employee))
    \end{equation*}
\end{demo}

\subsection*{Pipeline Notation}

Queries are often constructed incrementally, by extracting relevant data and
then shaping it into the desired form with a chain of combinators.  This
construction is made apparent with the \emph{pipeline notation}.

In pipeline notation, the first argument of a combinator is placed in front of
it, separated by colon (``$\,\Apply\,$''):
\begin{equation*}
    p \Apply F \equiv F(p), \qquad
    p \Apply F(q_1,\ldots,q_n) \equiv F(p,q_1,\ldots,q_n).
\end{equation*}
For example, $\Count(\Department)$ could also be written
\begin{equation*}
    \Department\Apply\Count.
\end{equation*}

A more sophisticated query written in pipeline notation is shown in the
following example.

\begin{demo}
    \label{ex:top-ten-highest-paid-policemen}
    Show the top 10 highest paid employees in the Police department.
    \begin{align*}
        & \Employee \\
        & \Apply\Filter(\Department\To\Name = \textliteral{POLICE}) \\
        & \Apply\Sort(\Salary\Apply\Desc) \\
        & \Apply\Select(\Name,\; \Position,\; \Salary) \\
        & \Apply\Take(10)
    \end{align*}
\end{demo}

Without pipeline notation, this query is much less intelligible:
\begin{multline*}
    \Take(\Select(\Sort(\Filter( \\
    \Employee,\; \Department\To\Name = \textliteral{POLICE}), \\
    \Desc(\Salary)),\; \Name,\; \Position,\; \Salary),\; 10).
\end{multline*}

The combinators $\Filter$, $\Sort$, $\Desc$, $\Select$, and $\Take$ are
described below.

\subsection*{Filtering Data}

We can now demonstrate how to produce entities that satisfy a certain
condition.

\begin{demo}
    \label{ex:filter-by-salary}
    Which employees have a salary higher than \$150k?
    \begin{equation*}
        \Employee\Apply\Filter(\Salary>150000)
    \end{equation*}
\end{demo}

This query introduces several concepts.

First, the integer literal $150000$ represents a primitive query that
\emph{for any given employee, produces the number $150000$}
\begin{equation*}
    150000 : \Emp \to \Int = e \mapsto 150000.
\end{equation*}

Second, the relational symbol ${>}$ denotes a binary combinator that builds a
query \emph{for a given employee, show whether their salary is higher than
\$150k}
\begin{equation*}
    \Salary > 150000 : \Emp \to \Bool.
\end{equation*}
The combinator
\begin{equation*}
    \placeholder>\placeholder : (A \to \Int,\; A \to \Int) \to (A \to \Bool)
\end{equation*}
is implemented by lifting the relational operator
\begin{equation*}
    \placeholder>\placeholder : (\Int,\; \Int) \to \Bool
\end{equation*}
to an operation on queries
\begin{equation*}
    (p > q) = a \mapsto (p(a) > q(a)).
\end{equation*}

Third, a binary combinator $\Filter$ emits those $\Employee$ entities that
satisfy the condition $\Salary > 150000$.  In general, given
\begin{equation*}
    p : A \to \Seq{B}, \qquad q : B \to \Bool,
\end{equation*}
a query
\begin{equation*}
    \Filter(p,\; q) : A \to \Seq{B}
\end{equation*}
produces the values of $p$ that satisfy condition $q$
\begin{equation*}
    \Filter(p,\; q) = a \mapsto [\,b \mid b \gets p(a),\; q(b)=\True\,].
\end{equation*}

The following example shows how $\Filter$ could be used in tandem with
aggregate combinators.

\begin{demo}
    \label{ex:filter-by-size-and-count}
    How many departments have more than 1000 employees?
    \begin{align*}
        & \Department \\
        & \Apply\Filter(\Count(\Employee)>1000) \\
        & \Apply\Count
    \end{align*}
\end{demo}

\subsection*{Sorting and Paginating Data}

The combinator $\Sort$, applied to a plural query, sorts the query output in
ascending order.

\begin{demo}
    \label{ex:sort-department-name}
    Show the names of all departments in alphabetical order.
    \begin{equation*}
        \Sort(\Department\To\Name)
    \end{equation*}
\end{demo}

The combinator $\Sort$ is implemented by lifting a sequence function
\begin{equation*}
    \Sort : \Seq{A} \to \Seq{A}
\end{equation*}
to a query combinator
\begin{align*}
    & \Sort : (A \to \Seq{B}) \to (A \to \Seq{B}), \\
    & \Sort(p) = a \mapsto \Sort(p(a)).
\end{align*}

\begin{demo}
    \label{ex:sort-employee-by-salary}
    Show all employees ordered by salary.
    \begin{equation*}
        \Employee\Apply\Sort(\Salary)
    \end{equation*}
\end{demo}

In this example, a list of employees is sorted by the value of the attribute
$\Salary$, which is supplied as the second argument to the $\Sort$ combinator.
In this form, $\Sort$ has a signature
\begin{equation*}
    \Sort : (A \to \Seq{B},\; B \to C) \to (A \to \Seq{B}).
\end{equation*}

\begin{demo}
    \label{ex:sort-employee-by-salary-desc}
    Show all employees ordered by salary, highest paid first.
    \begin{equation*}
        \Employee\Apply\Sort(\Salary\Apply\Desc)
    \end{equation*}
\end{demo}

Here, the sort key is wrapped with the combinator $\Desc$ to indicate the
descending sort order.

It is not immediately obvious how to implement $\Desc$ without violating the
query model.  Na\"{\i}vely, $\Desc$ acts like a negation operator, however, not
every type supports negation.  Instead, we make the sort order a part of the
type definition, so that
\begin{equation*}
    \Int_{\le} \quad\text{and}\quad \Int_{\ge}
\end{equation*}
could indicate the integer type with ascending and descending sort order
respectively.  Then, $\Desc$ could be considered a type conversion combinator
with the signature
\begin{equation*}
    \Desc : (A \to B) \to (A \to B_{\ge}).
\end{equation*}

\begin{demo}
    \label{ex:sort-employee-by-salary-take-top}
    Who are the top 1\% of the highest paid employees?
    \begin{align*}
        & \Employee \\
        & \Apply\Sort(\Salary\Apply\Desc) \\
        & \Apply\Take(\Count(\Employee)\mathbin \div 100)
    \end{align*}
\end{demo}

In this example, only the first 1\% of employees are retained by the combinator
$\Take$, which has two arguments: a query that produces a sequence of employees
\begin{equation*}
    \Employee\Apply\Sort(\Salary\Apply\Desc) : \Void \to \Seq{\Emp}
\end{equation*}
and a query that returns how many employees to keep
\begin{equation*}
    \Count(\Employee) \div 100 : \Void \to \Int.
\end{equation*}
Notice that both arguments of $\Take$ have the same input ($\Void$ in this
case), which is reflected in the signature
\begin{equation*}
    \Take : (A \to \Seq{B},\; A \to \Int) \to (A \to \Seq{B}).
\end{equation*}

\subsection*{Query Output}

The combinator $\Select$ customizes the query output.

Previously, we constructed a query to \emph{show the number of employees for
each department} (see Example~\ref{ex:department-count-employee}):
\begin{equation*}
    \Department\To\Count(\Employee).
\end{equation*}
However, this query only produces a list of bare numbers---it does not connect
them to their respective departments.  This is corrected in the following example.

\begin{demo}
    \label{ex:department-select-name-size}
    For each department, show its name and the number of employees.
    \begin{equation*}
        \Department
        \Apply\Select(\Name,\;\Size\As\Count(\Employee))
    \end{equation*}
\end{demo}

In this example, the combinator $\Select$ takes three arguments: the base query
\begin{equation*}
    \Department : \Void \to \Seq{\Dept}
\end{equation*}
and two field queries
\begin{alignat*}{3}
    & \Name && : \Dept && \to \Text, \\
    & \Count(\Employee) && : \Dept && \to \Int.
\end{alignat*}
The $\Select$ combinator generates a sequence of records by applying each field
query to every entity produced by the base query, giving this example a
signature
\begin{equation*}
    \Void \to \Seq{\Tuple{\Name:\Text,\; \Size:\Int}}.
\end{equation*}
The declaration
\begin{equation*}
    \Tuple{\Name:\Text,\; \Size:\Int}
\end{equation*}
defines a \emph{record} type with two fields: a text field $\Name$ and an
integer field $\Size$.  The names of the record fields are derived from the
tags of the field queries, which could be set using the \emph{tagging
notation}.  For example,
\begin{equation*}
    \Size \As \Count(\Employee)
\end{equation*}
binds a tag $\Size$ to the query $\Count(\Employee)$.  Since the tag does not
materially affect the query it annotates, we do not expose the tag in the query
model.

A more complex output structure could be defined with nested $\Select$ combinators.

\begin{demo}
    \label{ex:department-select-name-etc}
    For every department, show the top salary and a list of managers with their
    salaries.
    \begin{alignat*}{4}
        & \Department\hidewidth && && && \\
        & \Apply\Select( && \Name, && && \\
            & && \TopSalary && \As\; && \Max(\Employee\To\Salary), \\
            & && \Manager && \As\; && \Employee \\
            & && && && \Apply\Filter(\Exists(\Subordinate)) \\
            & && && && \Apply\Select(\Name,\; \Salary))
    \end{alignat*}
\end{demo}

In this example, the query output has the type
\begin{align*}
    \keyword{Seq}\{\Tuple{& \Name : \Text,\; \TopSalary : \Opt{\Int}, \\
    & \Manager : \Seq{\Tuple{\Name : \Text,\; \Salary : \Int}}}\}.
\end{align*}

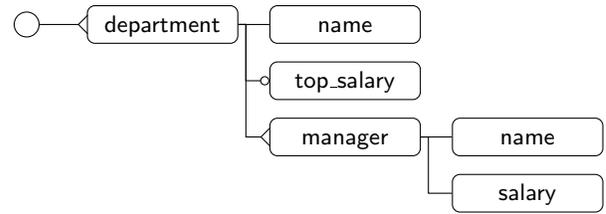
\begin{figure}
    \centering
    \begin{forest}
        unfolded database
        [,void
            [$\Department$,map,plural
                [$\Name$,map,singular]
                [$\TopSalary$,map,optional]
                [$\Manager$,map,plural
                    [$\Name$,map,singular],
                    [$\Salary$,map,singular]]]]
    \end{forest}
    \caption{Output database for Example~\ref{ex:department-select-name-etc}}
    \label{fig:department-select-name-etc}
\end{figure}

Recall that we represented the data source in a universal hierarchical form
(see Figure~\ref{fig:unfolded-form}).  Furthermore, the query output could also
be represented as a hierarchical database, whose structure is determined by the
query signature (see Figure~\ref{fig:department-select-name-etc}).  Thus,
queries could be seen as transformations of hierarchical databases.

\subsection*{Query Aliases}

A complex query could often be simplified by replacing duplicate expressions
with aliases.

\begin{demo}
    \label{ex:department-define-size}
    Show the top 3 largest departments and their sizes.
    \begin{align*}
        & \Department \\
        & \Apply\Define(\Size\As\Count(\Employee)) \\
        & \Apply\Sort(\Size\Apply\Desc) \\
        & \Apply\Select(\Name,\; \Size) \\
        & \Apply\Take(3)
    \end{align*}
\end{demo}

In this example, the alias $\Size$ is created in two steps: first, the tag
$\Size$ is bound to the query
\begin{equation*}
    \Count(\Employee): \Dept \to \Int,
\end{equation*}
and then $\Size$ is added to scope of $\Dept$ by the combinator $\Define$.

Although this query could have been written as
\begin{align*}
    & \Department \\
    & \Apply\Sort(\Count(\Employee)\Apply\Desc) \\
    & \Apply\Select(\Name,\; \Count(\Employee)) \\
    & \Apply\Take(3),
\end{align*}
the use of an alias makes this example more legible, not only by reducing
redundancy, but also by assigning a name to a key concept of the query.

\subsection*{Hierarchical Relationships}

Hierarchical relationships are encoded by self-referential primitives.

For example, the relationship between an employee and their manager is
expressed with
\begin{equation*}
    \Manager : \Emp \to \Opt{\Emp}.
\end{equation*}

\begin{demo}
    \label{ex:employee-filter-salary-manager}
    Find all employees whose salary is higher than the salary of their manager.
    \begin{equation*}
        \Employee\Apply\Filter(\Salary>\Manager\To\Salary)
    \end{equation*}
\end{demo}

This example uses familiar combinators $\Filter$ and ${>}$ (see
Example~\ref{ex:filter-by-salary}), but an alert reader will notice the
disagreement between the signature of the combinator
\begin{equation*}
    \placeholder>\placeholder : (A \to \Int,\; A \to \Int) \to (A \to \Bool)
\end{equation*}
and the signatures of its arguments
\begin{alignat*}{3}
    & \Salary && : \Emp && \to \Int, \\
    & \Manager\To\Salary && : \Emp && \to \Opt{\Int}.
\end{alignat*}
Namely, ${>}$ expects its arguments to be singular, but the output of
$\Manager\To\Salary$ is optional.

To legitimize this query, we adopt the following rule.  When one argument of a
scalar combinator has a non-trivial cardinality, this cardinality can be
promoted to the output of the combinator.  This rule gives ${>}$ a signature
\begin{equation*}
    \placeholder>\placeholder : (A \to \Int,\; A \to M\{\Int\}) \to (A \to M\{\Bool\})
\end{equation*}
or, in this specific case,
\begin{equation*}
    \Salary > \Manager\To\Salary : \Emp \to \Opt{\Bool}.
\end{equation*}
Finally, we need to let $\Filter$ accept predicate queries with optional
output, by treating $\bot$ as $\False$.

Using expressions
\begin{align*}
    & \Manager, \\
    & \Manager\To\Manager, \\
    & \Manager\To\Manager\To\Manager,\; \ldots
\end{align*}
we can build queries that involve the manager, the manager's manager, etc.  We
can also obtain \emph{the complete management chain for the given employee}
with
\begin{equation*}
    \Connect(\Manager) : \Emp \to \Seq{\Emp}.
\end{equation*}

\begin{demo}
    \label{ex:city-treasurer-subordinates}
    Find all direct and indirect subordinates of the City Treasurer.
    \begin{multline*}
        \Employee \\
        \shoveleft{\Apply\Filter(\Any(\Connect(\Manager)\To\Position =} \\
        \textliteral{CITY\ TREASURER}))
    \end{multline*}
\end{demo}

Here, the query
\begin{equation*}
    \Connect(\Manager)\To\Position : \Emp \to \Seq{\Text}
\end{equation*}
produces \emph{the positions of all managers above the given employee}.

In general, the combinator $\Connect$ maps an optional self-referential query to a
plural self-referential query by taking its transitive closure:
\begin{multline*}
    \Connect : (A \to \Opt{A}) \to (A \to \Seq{A}), \\
    \shoveleft{\Connect(p) = a \mapsto [\;p(a),\; p(p(a)),\; \ldots,\; p^{(n)}(a)\;]} \\
    (p^{(n)}(a) \ne \bot,\; p^{(n+1)}(a) = \bot).
\end{multline*}

\section{Quotient Classes}
\label{sec:quotients}

Previously, we demonstrated how to group and aggregate data---so long as the
structure of the data reflects the hierarchical form of the database.  In this
section, we show how to overcome this limitation.

In Figure~\ref{fig:unfolded-form}, the schema graph is unfolded into an
infinite tree, shaping the data into a hierarchical form.  A section of this
hierarchy could be extracted using the $\Select$ combinator.

\begin{demo}
    \label{ex:department-select-name-employee}
    Show all departments, and, for each department, list the associated
    employees.
    \begin{equation*}
        \Department\Apply\Select(\Name,\; \Employee)
    \end{equation*}
\end{demo}
But what if we ask for \emph{positions} instead of \emph{departments}?

\begin{demo}
    \label{ex:employee-group-position}
    Show all positions, and, for each position, list the associated employees.
\end{demo}

Unlike the previous example, this query does not match the structure of the
database and, therefore, cannot be constructed as easily.  Indeed,
Example~\ref{ex:department-select-name-employee} is built from the primitives
\begin{alignat*}{3}
    & \Department && : \Void && \to \Seq{\Dept}, \\
    & \Name && : \Dept && \to \Text, \\
    & \Employee && : \Dept && \to \Seq{\Emp}.
\end{alignat*}
To construct Example~\ref{ex:employee-group-position} in a similar fashion, we
need a hypothetical class $\Pos$ of \emph{position} entities and a set of
queries with the corresponding signatures
\begin{equation}
    \label{eq:group-components}
    \begin{alignedat}{2}
        & \Void && \to \Seq{\Pos}, \\
        & \Pos && \to \Text, \\
        & \Pos && \to \Seq{\Emp}.
    \end{alignedat} \tag{$\star\star\star$}
\end{equation}
However, there is no built-in class of position entities and we only have the
following primitives available:
\begin{alignat*}{3}
    & \Employee && : \Void && \to \Seq{\Emp}, \\
    & \Position && : \Emp && \to \Text.
\end{alignat*}

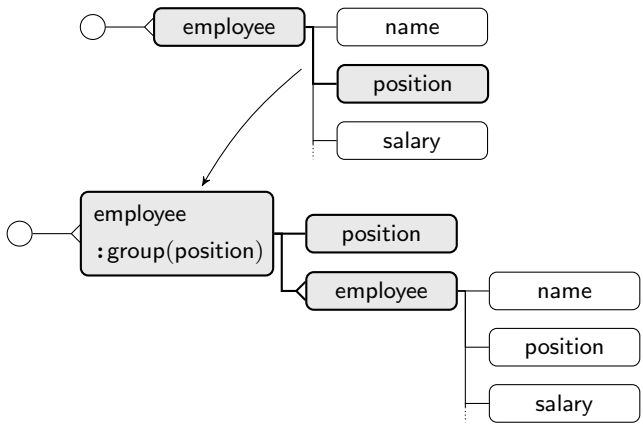
\begin{figure}
    \centering
    \begin{forest}
        unfolded database
        [,phantom
            [,phantom
                [,void
                    [$\Employee$,map,plural,selected,name=src employee
                        [$\Name$,map,singular]
                        [$\Position$,map,singular,selected,selected edge]
                        [$\Salary$,map,singular,and more]
                        [,phantom]]]]
            [,void
                [$\begin{aligned}&\Employee\\&\Apply\Group(\Position)\end{aligned}$,box,plural,selected,name=tgt group
                    [$\Position$,map,singular,selected,selected edge]
                    [$\Employee$,map,plural,selected,selected edge
                        [$\Name$,map,singular]
                        [$\Position$,map,singular]
                        [$\Salary$,map,singular,and more]]]]]
        \draw[traverse] (src employee.south east) +(0,-0.25cm) to [bend right=10] (tgt group);
    \end{forest}
    \caption{Action of the $\Group$ combinator}
    \label{fig:group-combinator}
\end{figure}

To make a ``virtual'' entity class from all distinct values of an attribute and
inject this class into the database structure, we use the $\Group$ combinator.
For example (see Figure~\ref{fig:group-combinator}), a list of \emph{all
distinct employee positions} can be produced with the query
\begin{equation*}
    \Employee\Apply\Group(\Position) : \Void \to \Seq{\Pos}.
\end{equation*}
The virtual $\Pos$ class comes with the primitives
\begin{alignat*}{3}
    & \Position && : \Pos && \to \Text, \\
    & \Employee && : \Pos && \to \Seq{\Emp},
\end{alignat*}
which, \emph{given a position entity, produce respectively the position name
and a list of associated employees}.
This gives us all the query components (see~(\ref{eq:group-components}) above)
needed to complete the example.

\addtocounter{demo}{-1}
\begin{demo}
    Show all positions, and, for each position, list the associated employees.
    \begin{align*}
        & \Employee \\
        & \Apply\Group(\Position) \\
        & \Apply\Select(\Position,\; \Employee)
    \end{align*}
\end{demo}

The query
\begin{equation*}
    \Employee\Apply\Group(\Position)
\end{equation*}
correlates all distinct values emitted by $\Position$ with the respective
$\Employee$ entities and packs them together into the records of type
\begin{equation*}
    \Pos \equiv \Tuple{\Position : \Text,\; \Employee : \Seq{\Emp}}.
\end{equation*}
We call $\Pos$ a \emph{quotient class} and denote it by
\begin{equation*}
    \Quotient{\Emp}{\Position}.
\end{equation*}

Once the database hierarchy is rearranged to include the class $\Pos$, we can
answer any questions about position entities.

\begin{demo}
    \label{ex:employee-group-position-etc}
    In the Police department, show all positions with the number of employees
    and the top salary.
    \begin{alignat*}{2}
        & \Employee\hidewidth && \\
        & \Apply\Filter(\Department\To\Name = \textliteral{POLICE})\hspace{-15em}&&\hspace{12em} \\
        & \Apply\Group(\Position)\hidewidth && \\
        & \Apply\Select(&& \Position, \\
        & && \Count(\Employee), \\
        & && \Max(\Employee\To\Salary))
    \end{alignat*}
\end{demo}

Here, for each position in the Police department, we determine two calculated
attributes, the number of employees and the top salary:
\begin{alignat*}{3}
    & \Count(\Employee) && : \Quotient{\Emp}{\Position} && \to \Int, \\
    & \Max(\Employee\To\Salary) && : \Quotient{\Emp}{\Position} && \to \Opt{\Int}.
\end{alignat*}

\begin{demo}
    \label{ex:nested-group}
    Arrange employees into a hierarchy: first by position, then by department.
    \begin{alignat*}{2}
        & \Employee\hidewidth && \\
        & \Apply\Group(\Position)\hidewidth && \\
        & \Apply\Select(&& \Position, \\
        & && \Employee \\
        & && \Apply\Group(\Department) \\
        & && \Apply\Select(\Department\To\Name,\; \Employee))
    \end{alignat*}
\end{demo}

Nested $\Group$ combinators can construct a hierarchical output of an arbitrary
form.  In this example, we rebuild the database hierarchy to place positions on
top, then departments, and then employees.  Notably, the nested $\Group$
expression has a signature
\begin{multline*}
    \Employee\Apply\Group(\Department) : \\
    \Quotient{\Emp}{\Position} \to \Seq{\Quotient{\Emp}{\Department}}.
\end{multline*}

\begin{demo}
    \label{ex:unique-department}
    Show all positions available in more than one department, and, for each
    position, list the respective departments.
    \begin{alignat*}{2}
        & \Employee\hidewidth &&  \\
        & \Apply\Group(\Position)\hidewidth && \\
        & \Apply\Define(&& \Department\As \\
        & && \qquad \Unique(\Employee\To\Department)) \\
        & \Apply\Filter(\Count(\Department) > 1)\hidewidth && \\
        & \Apply\Select(\Position,\; \Department\To\Name)\hidewidth &&
    \end{alignat*}
\end{demo}

This example uses the $\Unique$ combinator to find all distinct entities in a
list of departments.  The $\Unique$ combinator can be expressed via $\Group$ by
forgetting the plural component of the quotient class.  In this example,
$\Unique(\Employee\To\allowbreak\Department)$ is equivalent to
\begin{equation*}
    \Employee\Apply\Group(\Department)\To\Department.
\end{equation*}

\begin{demo}
    \label{ex:employee-group-level}
    How many employees at each level of the organization chart?
    \begin{align*}
        & \Employee \\
        & \Apply\Group(\Level\As\Count(\Connect(\Manager))) \\
        & \Apply\Select(\Level,\; \Count(\Employee))
    \end{align*}
\end{demo}

In order to apply $\Group$ to a calculated attribute, such as \emph{the level
in the organization chart}
\begin{equation*}
    \Count(\Connect(\Manager)) : \Emp \to \Int,
\end{equation*}
we need to bind an explicit tag to this attribute.

\begin{demo}
    \label{ex:rollup}
    Show the average salary by department and position, with subtotals for each
    department and the grand total.
    \begin{alignat*}{2}
        & \Employee\hidewidth && \\
        & \Apply\Rollup(\Department,\; \Position)\hidewidth && \\
        & \Apply\Select(&& \Department, \\
        & && \Position, \\
        & && \Mean(\Employee\To\Salary))
    \end{alignat*}
\end{demo}

To summarize data along several dimensions, we can apply $\Group$ to more than
one attribute.  When the summary data has to include subtotals and totals, we
replace $\Group$ with $\Rollup$.

In this example, the query
\begin{equation*}
    \Employee\Apply\Rollup(\Department,\; \Position)
\end{equation*}
produces a sequence of records of type
\begin{multline*}
    \Quotient{\Emp}{\Department_{\bot},\Position_{\bot}} \equiv \\
    \begin{alignedat}{3}
        & \Tuple{&& \Department && : \Opt{\Dept}, \\
        & && \Position && : \Opt{\Text}, \\
        & && \Employee && : \Seq{\Emp}}.
    \end{alignedat}
\end{multline*}
In addition to the records that would be generated by $\Group$, $\Rollup$ emits
one ``subtotal'' record per each department and one ``grand total'' record.
The former has the $\Position$ field set to $\bot$ and an $\Employee$ list
containing all employees in the given department.  The latter has both
$\Department$ and $\Position$ set to $\bot$ and $\Employee$ containing the full
list of employees.

\section{Query Context}
\label{sec:context}

In this section, we extend the query model to support \emph{context-aware}
queries: parameterized queries and queries with window functions.

\begin{demo}
    \label{ex:parameters}
    Show all employees in the given department $\DEPARTMENT$ with the salary
    higher than $\SALARY$, where
    \begin{equation*}
        \DEPARTMENT=\textliteral{POLICE}, \quad \SALARY=150000.
    \end{equation*}
    \begin{align*}
        & \Employee \\
        & \Apply\Filter(\Department\To\Name = \DEPARTMENT \mathbin{\&} \Salary > \SALARY) \\
        & \Apply\Given(\DEPARTMENT \As \textliteral{POLICE},\; \SALARY \As 150000)
    \end{align*}
\end{demo}

Practical database queries often depend upon \emph{query parameters}, which
collectively form the query \emph{environment}.  The environment is represented
by a container, such as
\begin{equation*}
    \Env{\DEPARTMENT:\Text,\SALARY:\Int}{A} \equiv \Tuple{A,\; \Tuple{\DEPARTMENT:\Text,\; \SALARY:\Int}},
\end{equation*}
that encapsulates both the regular input value and the values of the
parameters.  The parameters can be extracted from the environment with the
primitives
\begin{equation*}
    \DEPARTMENT : \Env{\DEPARTMENT:\Text}{A} \to \Text, \qquad
    \SALARY : \Env{\SALARY:\Int}{A} \to \Int.
\end{equation*}

The query environment is populated using the combinator $\Given$.  In this
example, the first argument of $\Given$ is a parameterized query
\begin{multline*}
    \Employee \\
    \shoveleft{\Apply\Filter(\Department\To\Name = \DEPARTMENT \mathbin{\&} \Salary > \SALARY) : } \\
    \Env{\DEPARTMENT:\Text,\SALARY:\Int}{\Void} \to \Seq{\Emp}.
\end{multline*}
The other two arguments are the constant queries
\begin{equation*}
    \textliteral{POLICE} : \Void \to \Text, \qquad
    150000 : \Void \to \Int
\end{equation*}
that specify the values of the parameters.  The combined query does \emph{not}
depend upon the parameters, and, hence, has a signature
\begin{equation*}
    \Void \to \Seq{\Emp}.
\end{equation*}

In general, $\Given$ takes a parameterized query
\begin{equation*}
    p : \Env{x_1:T_1,\ldots,x_n:T_n}{A} \to \Wrap{M}{B},
\end{equation*}
$n$ queries that evaluate the parameters
\begin{equation*}
    q_1 : A \to T_1,\quad \ldots,\quad q_n : A \to T_n
\end{equation*}
and combines them into a context-free query
\begin{align*}
    & \Given(p,\; q_1,\; \ldots,\; q_n) : A \to \Wrap{M}{B}, \\
    & \Given(p,\; q_1,\; \ldots,\; q_n) = a \mapsto p(\Tuple{a, \Tuple{q_1(a), \ldots, q_n(a)}}).
\end{align*}

\begin{demo}
    \label{ex:higher-than-average-salary}
    Which employees have higher than average salary?
    \begin{align*}
        & \Employee \\
        & \Apply\Filter(\Salary > \MEANSALARY) \\
        & \Apply\Given(\MEANSALARY \As \Mean(\Employee\To\Salary))
    \end{align*}
\end{demo}

This example uses the query environment to pass information between different
scopes.  The parameter $\MEANSALARY$ is calculated in the scope of $\Void$ by
the query
\begin{equation*}
    \Mean(\Employee\To\Salary) : \Void \to \Opt{\Num}
\end{equation*}
and is extracted in the scope of $\Emp$ by the primitive
\begin{equation*}
    \MEANSALARY : \Env{\MEANSALARY:\Opt{\Num}}{\Emp} \to \Opt{\Num}.
\end{equation*}

The query environment is one example of a \emph{query context}, a comonadic
container wrapping the query input.  It could be shown that the environment is
compatible with query composition (cf.~Section~\ref{sec:cardinality}), which
permits us to incorporate it into the query model.

Another example of a query context is the \emph{input flow}, a container of all
input values seen by the query.  We denote this context type by $\Rel{A}$ and
its values by
\begin{equation*}
    [a_1,\;\ldots,\selected{a_j},\;\ldots,\;a_n] : \Rel{A},
\end{equation*}
where $a_j$ is the current input value, $a_1,\;\ldots,\;a_{j-1}$ are the values
seen in the past, and $a_{j+1},\;\ldots,\;a_n$ are the values to be seen in the
future.  The input flow can be used for an alternative implementation of
Example~\ref{ex:higher-than-average-salary}.

\begin{demobis}{ex:higher-than-average-salary}
    Which employees have higher than average salary?
    \begin{equation*}
        \Employee\Apply\Filter(\Salary > \Mean(\Around\To\Salary))
    \end{equation*}
\end{demobis}

To relate each value in a dataset to the dataset as a whole, we use the plural
primitive $\Around$, which materializes its input flow as a sequence:
\begin{multline*}
    \Around : \Rel{A} \to \Seq{A} \\
    \shoveleft{\Around = [a_1,\;\ldots,\selected{a_j},\;\ldots,\;a_n]} \\
    \mapsto [a_1,\;\ldots,\;a_j,\;\ldots,\;a_n].
\end{multline*}
In this example, $\Around$ produces, \emph{for a selected employee, a list of
all employees}.  By composing it with $\Salary$, we get, \emph{for a selected
employee, a list of all salaries}
\begin{equation*}
    \Around\To\Salary : \Rel{\Emp} \to \Seq{\Int},
\end{equation*}
which lets us establish \emph{the average salary} as a context-aware attribute
\begin{equation*}
    \Mean(\Around\To\Salary) : \Rel{\Emp} \to \Opt{\Num}.
\end{equation*}

\begin{demo}
    \label{ex:higher-than-average-salary-by-position}
    In the Police department, show employees whose salary is higher than the
    average for their position.
    \begin{align*}
        & \Employee \\
        & \Apply\Filter(\Department\To\Name=\textliteral{POLICE}) \\
        & \Apply\Filter(\Salary > \Mean(\Around(\Position)\To\Salary))
    \end{align*}
\end{demo}

Here, each employee is matched with other employees having the same position
using a variant of $\Around$:
\begin{multline*}
    \Around : (A \to B) \to (\Rel{A} \to \Seq{A}) \\
    \shoveleft{\Around(q) = [a_1,\;\ldots,\selected{a_j},\;\ldots,\;a_n]} \\
    \mapsto [\,a_i \mid q(a_i)=q(a_j)\,].
\end{multline*}
Note the use of two separate $\Filter$ combinators.  If we switch them,
$\Around(\Position)$ would list employees with the same position \emph{across
all departments}.

We can exploit the input flow to calculate running aggregates.

\begin{demo}
    \label{ex:employee-select-no-total}
    Show a numbered list of employees and their salaries along with the running
    total.
    \begin{alignat*}{3}
        & \Employee\hidewidth && && \\
        & \Apply\Select(&& \No && \As \Count(\Before), \\
        & && \Name,\hidewidth && \\
        & && \Salary,\hidewidth &&  \\
        & && \Total && \As \Sum(\Before\To\Salary))
    \end{alignat*}
\end{demo}

The primitive $\Before$ exposes its input flow up to and including the current
input value:
\begin{align*}
    & \Before : \Rel{A} \to \Seq{A} \\
    & \Before = [a_1,\;\ldots,\selected{a_j},\;\ldots,\;a_n] \mapsto [a_1,\;\ldots,\;a_j].
\end{align*}
Using $\Before$, we can enumerate the rows in the output
\begin{equation*}
    \Count(\Before) : \Rel{\Emp} \to \Int
\end{equation*}
as well as calculate the running sum of salaries
\begin{equation*}
    \Sum(\Before\To\Salary) : \Rel{\Emp} \to \Int.
\end{equation*}

\begin{demo}
    \label{ex:department-frame-employee}
    For each department, show employee salaries along with the running total;
    the total should be reset at the department boundary.
    \begin{alignat*}{2}
        & \Department\hidewidth && \\
        & \Apply\Select(&& \Name, \\
        & && \Employee \\
        & && \Apply\Select(\Name,\; \Salary,\; \Sum(\Before\To\Salary)) \\
        & && \Apply\Frame)
    \end{alignat*}
\end{demo}

The input flow propagates through composition, so that a query executed within
the context of
\begin{equation*}
    \Department\To\Employee : \Void \to \Seq{\Emp}
\end{equation*}
will see the input flow containing all the employees in all departments.  To
reset the input flow at a certain boundary, we use the combinator
\begin{equation*}
    \Frame : (\Rel{A} \to \Wrap{M}{B}) \to (A \to \Wrap{M}{B}).
\end{equation*}

\section{Conclusion and Related Work}
\label{sec:conclusion}

In this paper, we introduce a combinator-based query language, \emph{Rabbit},
and, using the framework of (co)monads and (bi-)Kleisli arrows \cite{Moggi1991,
Uustalu2005}, describe the denotation of database queries.

The functional database model presents the database as a collection of
extensionally defined arrows in some underlying category of serializable data.
We bootstrap the query model by assuming that a query with the input of type
$A$ and the output of type $B$ can be expressed in this category as an arrow
\begin{equation*}
    A \to B.
\end{equation*}
To model optional and plural queries, we wrap their output in a monadic
container and represent them as Kleisli arrows
\begin{equation*}
    A \to \Wrap{M}{B}.
\end{equation*}
The containers should form a family $\mathcal{M}$ of monads equipped with a
join-semilattice structure: for any $M_1, M_2 \in \mathcal{M}$, there exists
$M_1 \sqcup M_2 \in \mathcal{M}$ with natural injections
\begin{equation*}
    \Wrap{M_1}{A} \rightarrow \Wrap{(M_1 \sqcup M_2)}{A} \leftarrow \Wrap{M_2}{A}.
\end{equation*}
To represent query parameters and the input flow, we wrap the query input in a
comonadic container, expressing context-aware queries as bi-Kleisli arrows
\begin{equation*}
    \Wrap{W}{A} \to \Wrap{M}{B}.
\end{equation*}
Dually, the comonadic containers form a meet-semi\-lat\-tice $\mathcal{W}$ of
comonads: for any $W_1, W_2 \in \mathcal{W}$, there exists
$W_1 \sqcap W_2 \in \mathcal{W}$ with natural projections
\begin{equation*}
    \Wrap{W_1}{A} \leftarrow (W_1 \sqcap W_2)\{A\} \rightarrow \Wrap{W_2}{A}.
\end{equation*}
Moreover, for any monad $M\in\mathcal{M}$ and comonad $W\in\mathcal{W}$, there
should exist a distributive law
\begin{equation*}
    \Wrap{W}{\Wrap{M}{A}} \to \Wrap{M}{\Wrap{W}{A}}.
\end{equation*}
Then, the composition of queries
\begin{equation*}
    p : \Wrap{W_1}{A} \to \Wrap{M_1}{B}, \quad q : \Wrap{W_2}{B} \to \Wrap{M_2}{C}
\end{equation*}
could be defined as a query of the form
\begin{multline*}
    p\,\To\,q : \Wrap{W}{A} \to \Wrap{M}{C} \\ (W = W_1 \sqcap W_2,\; M = M_1 \sqcup M_2)
\end{multline*}
constructed using the lattice structures of $\mathcal{M}$ and $\mathcal{W}$,
compositional properties of monads and comonads, and the distributive law for
$M$ and $W$.

Rabbit has its roots in the authors' work on a URL-based query
language~\cite{Evans2007}, which provided a navigational interface to SQL
databases.  While looking for a way to formally specify this language, we
arrived at the combinator-based query model.

Early on, we adopted the navigational approach of XPath~\cite{Clark1999}, which
led us to represent the schema as a rooted graph (e.g.,
Figure~\ref{fig:folded-form}) and queries as paths in this graph.  We
recognized that each graph arc has some cardinality, and, consequently, so does
each path.  Next came the realization that, for any dataset, the dataset values
are all related to each other, and this relationship can be denoted as a plural
self-referential arc $\Around$.  We discovered that the rule for composing
$\Around$ with other plural arcs is exactly the distributive law for the
$\type{Rel}$ comonad over the $\type{Seq}$ monad, which pushed us to model
database queries as Kleisli arrows.

Monads and their Kleisli arrows came to be a standard tool in denotational
semantics after Moggi~\cite{Moggi1991} used them to define a generic
compositional model of computations.  By varying the choice of monad, he
expressed partiality, exceptions, input-output, and other computational
effects.  Uustalu and Vene~\cite{Uustalu2005} used a dual model of comonads and
co-Kleisli arrows to describe semantics of dataflow programming.  They also
discussed distributive laws of a comonad over a monad.  In the context of
databases, Spivak~\cite{Spivak2012} suggested using monads to encode data with
complex structure.  Monad comprehensions~\cite{Trinder1989, Buneman1994} form
the core of query interfaces such as Kleisli~\cite{Wong2000} and
LINQ~\cite{Meijer2006}.  In contrast with Rabbit, which is based on Kleisli
arrows and monadic composition, these interfaces are designed around monadic
containers and the monadic \emph{bind} operator.

The graph representation of the database schema is a variation of the
functional database model~\cite{Kerschberg1976, Sibley1977}, which gave rise to
a number of query languages: FQL~\cite{Buneman1979}, DAPLEX~\cite{Shipman1981},
GENESIS~\cite{Batory1988}, Kleisli~\cite{Wong2000} and others;
see~\cite{Gray2004} for a comprehensive survey.  Among them, FQL and its
derivatives are remarkably close to Rabbit---Example~\ref{ex:composite-query}
is a valid query in both.  The key difference is that we interpret the period
(``$\To$'') as a composition of Kleisli arrows, which implies, for instance,
that we cannot define $\Count$ as $\Seq{A}\to\Int$ and write
$\Employee\To\Count$ for \emph{the number of employees}.  Instead, we have to
accept $\Count$ as a query combinator.

Combinators are higher-order functions that serve to construct expressions
without bound variables.  They were introduced as the building blocks of
mathematical logic~\cite{Schoenfinkel1924, Curry1930}, from where they migrated
to programming practice, becoming a popular tool for constructing DSLs;
examples are found in diverse domains such as parsers~\cite{Wadler1985,
Hutton1996}, reactive animation~\cite{Elliott1997}, financial
contracts~\cite{Jones2000}, and the view-update problem~\cite{Foster2005}.

Although a few combinator-based query models have been
proposed~\cite{Buneman1979, Bossi1984, Batory1988, Erwig1991, Cherniack1996},
it is generally accepted that ``combinator-style languages are difficult for
users to master and thus ill-suited as query languages''~\cite{Cherniack1996}.
Examples presented in this paper prove otherwise.  Moreover, the syntax of a
combinator-based DSL directly mirrors its semantics, making it an
\emph{executable specification}.  This is an attractive property for a language
oriented towards domain experts---if the semantics does not contradict the
experts' intuition.

In Rabbit, the intuition relies upon the hierarchical data model, which is
simple, familiar and prolific.  For querying purposes, we view the database as
a composite hierarchical data structure obtained by unfolding the database
schema into a potentially infinite schema tree (e.g.,
Figure~\ref{fig:unfolded-form}).  We were inspired by concurrency theory, where
static ``system'' models are unfolded into runtime ``behavior''
models~\cite{Nielsen1994}, but this technique has also been used in database
theory to relate the network and hierarchical data models~\cite{Cartmell1985}.

Rabbit's query model lets us rigorously define the basic notions of data
analysis.  Indeed, it can naturally express optional and plural relationships;
database navigation; transitive closure of hierarchical relationships;
aggregate, grouping and data cube operations; query parameters and window
functions.  In fact, any data operation could be lifted to a query combinator.

For specific application domains, Rabbit can provide an extensible query
framework.  Applications can implement native domain operations by extending
the sets of primitives, combinators, and (co)monadic containers.  For example,
we adapted Rabbit to the field of medical informatics by adding graph
operations over hierarchical ontologies and temporal operations on medical
observations.

For its users, Rabbit can provide a collaborative data processing platform.
Database queries should be seen as artifacts of informatics
collaboration---transparent, executable specifications that are written,
shared, and discussed by software developers, data analysts, statisticians, and
subject-matter experts.  We believe that a compositional query model focused on
data relationships can enable this dialog.

\section{Acknowledgements}

We are indebted to Catherine Devlin for her early support of the project, and
our colleagues at Prometheus Research for their continuous feedback.

\bibliographystyle{abbrv}
\bibliography{rbt-paper}

\begin{thebibliography}{10}

\bibitem{Batory1988}
D.~S. Batory, T.~Y. Leung, and T.~E. Wise.
\newblock Implementation concepts for an extensible data model and data
  language.
\newblock {\em {ACM} Transactions on Database Systems}, 13(3):231--262, 1988.

\bibitem{Bossi1984}
A.~Bossi and C.~Ghezzi.
\newblock Using {FP} as a query language for relational data-bases.
\newblock {\em Computer Languages}, 9(1):25--37, 1984.

\bibitem{Buneman1979}
P.~Buneman and R.~E. Frankel.
\newblock {FQL} --- {A} functional query language.
\newblock In {\em {SIGMOD} '79}, pages 52--58, 1979.

\bibitem{Buneman1994}
P.~Buneman, L.~Libkin, D.~Suciu, V.~Tannen, and L.~Wong.
\newblock Comprehension syntax.
\newblock {\em {SIGMOD} Record}, 23(1):87--96, 1994.

\bibitem{Cartmell1985}
J.~Cartmell.
\newblock Formalizing the network and hierarchical data models --- an
  application of categorical logic.
\newblock In {\em {CTCS} '85}, pages 466--492, 1985.

\bibitem{Cherniack1996}
M.~Cherniack and S.~B. Zdonik.
\newblock Rule languages and internal algebras for rule-based optimizers.
\newblock In {\em {SIGMOD} '96}, pages 401--412, 1996.

\bibitem{Clark1999}
J.~Clark and S.~DeRose.
\newblock {XML} path language ({XPath}) version 1.0.
\newblock Technical Report REC-xpath-19991116, W3C, 1999.

\bibitem{Curry1930}
H.~B. Curry.
\newblock Grundlagen der {K}ombinatorischen {L}ogik.
\newblock {\em American Journal of Mathematics}, 52(3):509--536, 1930.

\bibitem{Elliott1997}
C.~Elliott and P.~Hudak.
\newblock Functional reactive animation.
\newblock In {\em {ICFP} '97}, pages 263--273, 1997.

\bibitem{Erwig1991}
M.~Erwig and U.~W. Lipeck.
\newblock A functional {DBPL} revealing high level optimizations.
\newblock In {\em {DBPL} '91}, pages 306--321, 1991.

\bibitem{Evans2007}
C.~C. Evans.
\newblock {HTSQL} --- a native web query language.
\newblock In {\em {ICOMP} '07}, pages 439--445, 2007.

\bibitem{Foster2005}
J.~N. Foster, M.~B. Greenwald, J.~T. Moore, B.~C. Pierce, and A.~Schmitt.
\newblock Combinators for bi-directional tree transformations: a linguistic
  approach to the view update problem.
\newblock In {\em {POPL} '05}, pages 233--246, 2005.

\bibitem{Gray2004}
P.~M.~D. Gray, P.~J.~H. King, and A.~Poulovassilis.
\newblock Introduction to the use of functions in the management of data.
\newblock In P.~M.~D. Gray, L.~Kerschberg, P.~J.~H. King, and A.~Poulovassilis,
  editors, {\em The Functional Approach to Data Management}, pages 1--54.
  Springer, Berlin, Heidelberg, 2004.

\bibitem{Hutton1996}
G.~Hutton and E.~Meijer.
\newblock Monadic parser combinators.
\newblock Technical Report NOTTCS-TR-96-4, School of Computer Science and IT,
  University of Nottingham, 1996.

\bibitem{Jones2000}
S.~L.~P. Jones, J.~Eber, and J.~Seward.
\newblock Composing contracts: an adventure in financial engineering.
\newblock In {\em {ICFP} '00}, pages 280--292, 2000.

\bibitem{Kerschberg1976}
L.~Kerschberg and J.~E.~S. Pacheco.
\newblock A functional data base model.
\newblock Technical Report 2/1976, Departamento de Informatica, Pontificia
  Universidade Catolica, Rio de Janeiro, Brazil, 1976.

\bibitem{Meijer2006}
E.~Meijer, B.~Beckman, and G.~M. Bierman.
\newblock {LINQ:} reconciling object, relations and {XML} in the {.NET}
  framework.
\newblock In {\em {SIGMOD} '06}, page 706, 2006.

\bibitem{Moggi1991}
E.~Moggi.
\newblock Notions of computation and monads.
\newblock {\em Information and Computation}, 93(1):55--92, 1991.

\bibitem{Nielsen1994}
M.~Nielsen, V.~Sassone, and G.~Winskel.
\newblock Relationships between models of concurrency.
\newblock In {\em {REX} '93}, pages 425--476, 1994.

\bibitem{Schoenfinkel1924}
M.~Sch{\"o}nfinkel.
\newblock {\"U}ber die {B}austeine der mathematischen {L}ogik.
\newblock {\em Mathematische Annalen}, 92(3):305--316, 1924.

\bibitem{Shipman1981}
D.~W. Shipman.
\newblock The functional data model and the data language {DAPLEX}.
\newblock {\em {ACM} Transactions on Database Systems}, 6(1):140--173, 1981.

\bibitem{Sibley1977}
E.~H. Sibley and L.~Kerschberg.
\newblock Data architecture and data model considerations.
\newblock In {\em {AFIPS} '77}, pages 85--96, 1977.

\bibitem{Spivak2012}
D.~I. Spivak.
\newblock Kleisli database instances.
\newblock {\em CoRR}, abs/1209.1011, 2012.

\bibitem{Trinder1989}
P.~Trinder and P.~Wadler.
\newblock Improving list comprehension database queries.
\newblock In {\em TENCON '89}, pages 186--192, 1989.

\bibitem{Uustalu2005}
T.~Uustalu and V.~Vene.
\newblock The essence of dataflow programming.
\newblock In {\em {CEFP} '05}, pages 135--167, 2005.

\bibitem{Wadler1985}
P.~Wadler.
\newblock How to replace failure by a list of successes.
\newblock In {\em {FPCA} '85}, pages 113--128, 1985.

\bibitem{Wong2000}
L.~Wong.
\newblock Kleisli, a functional query system.
\newblock {\em Journal of Functional Programming}, 10(1):19--56, 2000.

\end{thebibliography}

\end{document}